\documentclass{iau}

\usepackage{amsmath}
\usepackage{graphicx}
\usepackage{multirow}
\usepackage{natbib}

\begin{document}

\lefttitle{Gopalswamy {\it et al.}}
\righttitle{CME Catalog v2}

\jnlPage{1}{7}
\jnlDoiYr{2024}
\doival{10.1017/xxxxx}
\volno{388}
\pubYr{2024}
\journaltitle{Solar and Stellar Coronal Mass Ejections}

\aopheadtitle{Proceedings of the IAU Symposium}
\editors{N. Gopalswamy,  O. Malandraki, A. Vidotto \&  W. Manchester, eds.}

\title{The SOHO LASCO CME Catalog - Version 2}

\author{Nat Gopalswamy$^1$, Grzegorz Micha\l{}ek$^{2}$, Seiji Yashiro$^{3,1}$, Pertti M{\"a}kel{\"a}$^{3,1}$, Sachiko Akiyama$^{3,1}$, Hong Xie$^{3,1}$ and Angelos Vourlidas$^{4}$}
\affiliation{{}$^1$NASA Goddard Space Flight Center, Greenbelt, MD, USA\\email: \email{nat.gopalswamy@nasa.gov}}
\affiliation{{}$^2$Jagiellonian University, Krak{\'o}w, Poland}
\affiliation{{}$^3$The Catholic University of America, Washington, DC, USA}
\affiliation{{}$^4$Johns Hopkins University Applied Physics Laboratory, Laurel, MD, USA}

\begin{abstract}
This paper provides an update on the coronal mass ejection (CME) catalog maintained at the CDAW Data Center, NASA Goddard Space Flight Center (\url{http://cdaw.gsfc.nasa.gov/CME_list}).  This is version 2 (v2) of the Catalog that has been made as the default version as of May 1, 2024. The new features of the Catalog v2 are (i) online measurement tool, (ii) combination JavaScript movies from the STEREO and Solar Dynamics Observatory (SDO) missions, and (iii) insertion of newly identified CMEs for the period 1996 to 2004. The CME identification was revisited resulting in a set of $\sim$3000 new CMEs added to the Catalog. A vast majority of these CMEs are weak and narrow. The resulting statistical properties of CMEs are not significantly different from those reported using version 1.
\end{abstract}

\begin{keywords}
coronal mass ejections
\end{keywords}

\maketitle

\section{Introduction}

The CDAW CME Catalog grew out of a coordinated data analysis workshop (CDAW) in 1999 organized to study the properties of CMEs associated with interplanetary radio bursts. During the course of analyzing these events, it was realized that creating a set of plots, movies, and tables for each CME is a convenient way of investigating CME-related phenomena. The CMEs are manually identified in the images obtained by the Large Angle and Spectrometric Coronagraph \citep[LASCO,][]{1995SoPh..162..357B} on board the Solar and Heliospheric Observatory (SOHO). The Catalog resides in the CDAW Data Center at the Goddard Space Flight Center (\url{http://cdaw.gsfc.nasa.gov}) and is open to the scientific community. The version 1 catalog was initially reported in \citet{2004JGRA..109.7105Y}, which was expanded with new features as reported in \citet{2009EM&P..104..295G}. The number of CMEs listed in the CDAW catalog is an order of magnitude larger than the total number of CMEs known from pre-SOHO coronagraphs \citep{2009EM&P..104..295G,2023FrASS..1064226H}. This catalog serves as an important reference to test automatic detection catalogs such as cactus (\url{https://www.sidc.be/cactus/}) and SEEDS (\url{http://spaceweather.gmu.edu/seeds/}).

\section{Catalog}

\begin{figure}[]
  \centering
    \includegraphics[scale=1.0]{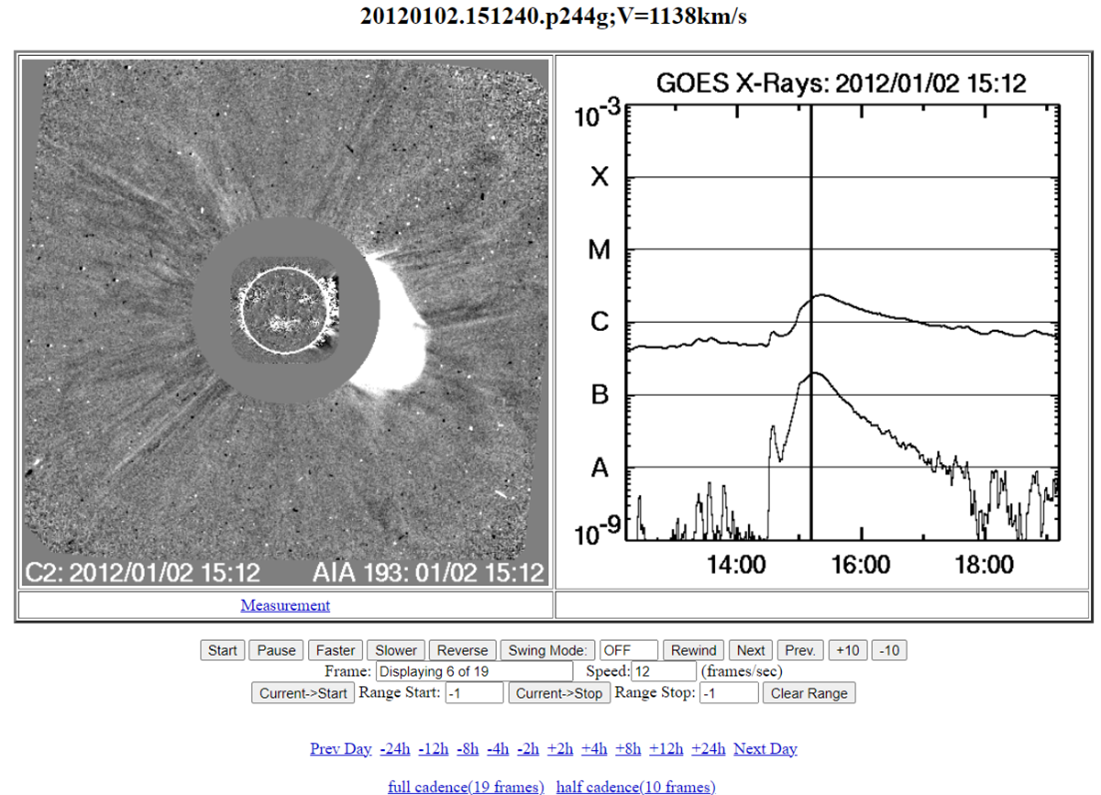}
    \caption{One frame in the JavaScript movie of the 2012 January 2 CME that first appeared in LASCO/C2 FOV at 15:12:40 UT. The movie is built using LASCO/C2 difference images with superposed EUV images from the Solar Dynamics Observatory (SDO, \citet{2012SoPh..275....3P}). The CME ID is given at the top as 20120102.151240.p244g. The numbers following p (for position angle) indicate the central position angle, while the suffix g is the first letter of the first name of the person who made the height-time measurement.  Also given at the top is the CME speed (V=1138 km s$^{-1}$) obtained from a linear fit to the height-time measurements. The JavaScript movie can be controlled using the buttons given below the movie frame. The length of the movie can be extended to earlier or later times up to 24 hours. The movie can also be run with full or half cadence. The link "Measurement" takes to a new page, where h-t measurements can be made.}
    \label{fig1}
  \end{figure}

The main feature of the catalog is a 13-column html table (\url{http://cdaw.gsfc.nasa.gov/CME_list/index.html}). The first column gives the year of CME observations; the next 12 columns give links to monthly list of CMEs. The monthly list is also a html table with each row giving all available information for one CME. The information is in the form of movies, plots, digital data, and remarks. The first three columns provide a unique identification (ID) for each CME: date (YYYY/MM/DD), universal time (HH:MM:SS), and position angle (in degrees). The first column is linked to a JavaScript movie constructed from LASCO/C2 running difference images with superposed EUV difference images. Figure~\ref{fig1} shows one frame of the JavaScript movie. The unique ID for the CME is given above the movie frame, in the format YYYYMMDD.HHMMSS.pAAAf;V=xxxxkm/s (see Fig.~\ref{fig1}) where p indicates that the succeeding three numbers give the central position angle of the CME in degrees; f indicates the first letter of the first name of the person who made the height-time measurement; V is the average speed of the CME within the coronagraph field of view (FOV). While the date is linked to the Javascript movie noted above, the time is linked to the height-time (h-t) data points as measured. The remaining columns in the monthly table are as described in the previous version \citep{2009EM&P..104..295G}. 

\subsection{The measurement tool}

One of the new features of the catalog is the online tool using which the users can make their own measurements. The measurements can be made to by clicking on the "Measurement" link given directly below the images. The catalog gives routine h-t measurements made at the fastest-moving leading edge of a CME. These measurements do not distinguish whether the leading edge is a shock or flux rope. The user can measure sub features such as shock, flux rope, or prominence core of a CME. Measurements can be made at various position angles if needed.  The measurement tool for radio dynamic spectra is mainly used for measuring drift rates of radio bursts. Based on the cursor position, in the measurement tool displays the pixel coordinates, distance from the Sun center in solar radii, the position angle, and the heliographic coordinates. Thus, one can find the solar source locations of eruptions by clicking on the disturbances observed in EUV images. 
  \begin{figure}[h]
  \centering
    \includegraphics[scale=.65]{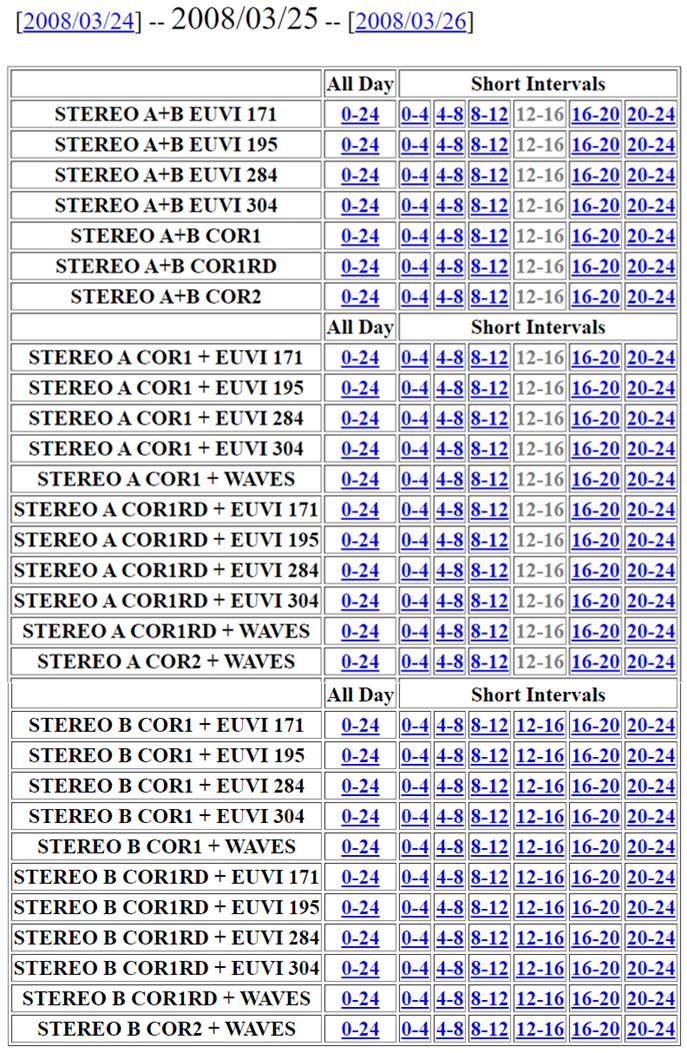}
    \caption{Table showing the set of new STEREO movies added to the catalog. The movies use various combinations of EUV images at 171, 195, 284, and 304 {\AA}, inner (COR1) and outer (COR2) coronagraph images, and WAVES dynamic spectra from STEREO Ahead (A) and Behind (B). Movies can be displayed for the whole day (0-24) or in 4-hour segments. The date of the movies is given at the top with options to display from the previous day or next day. Text in blue indicate links. Greyed out texts indicate data gaps.}
    \label{fig2}
  \end{figure}

The measurement tool for the radio dynamic spectra gives the time and frequency corresponding to the cursor position. Thus, by clicking on features in the dynamic spectra, one can get a table of time vs. frequency data, which can be used to derive the drift rates.  The radio dynamic spectra can be found under “Java Movie” in the column 12 of the catalog. For example, for the event in Fig.1, clicking on “Java Movie” opens the link,  \url{https://cdaw.gsfc.nasa.gov/CME_list/UNIVERSAL_ver2/2012_01/univ2012_01.html}. Under this link, one can find various combination movies including WAVES dynamic spectra. There are three movies that combine LASCO C2 and C3 images with Wind/WAVES dynamic spectra. Any of these can be used for making measurements on the dynamic spectra. A commonly used combination is “c2rdif\_waves.html”, which opens the movie,  \url{https://cdaw.gsfc.nasa.gov/movie/make_javamovie.php?date=20120102&img1=lasc2rdf&img2=wwaves}. In the movie, the left frames are running difference images (c2rdif) from SOHO/LASCO while the right frame is the WAVES dynamic spectrum. The measurement feature in this movie can be used for both CME h-t and drift rate measurements.  The java movie needs to be paused before clicking on the measurement tool at the WAVES frame containing the feature to be measured.

  \begin{figure}[h]
  \centering
    \includegraphics[scale=0.9]{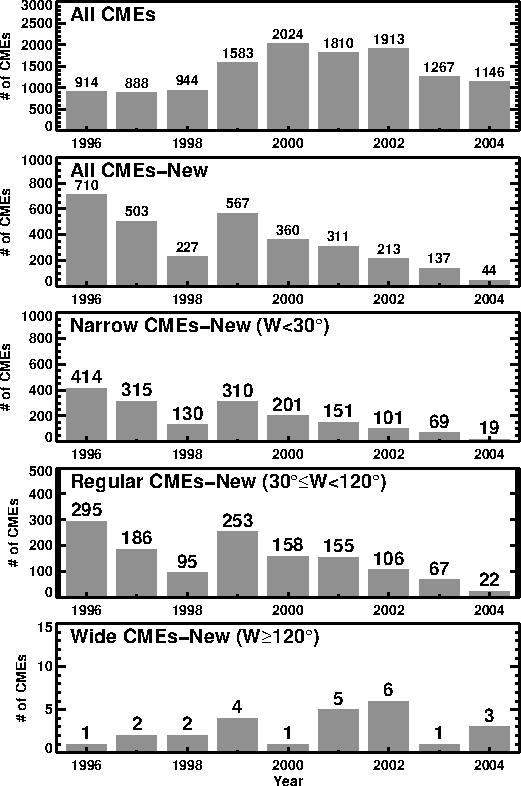}
    \caption{All CMEs in the catalog v2 over the period 1996 --2004 (top)  and the newly  added CMEs (lower 4 panels). The number of CMEs in each year are indicated on the bars of the plots. The only missing CME in the year 2000 was a faint halo CME that occurred on 2000 January 2. This is the only newly added halo CME for the entire interval (1996 to 2004). }
    \label{fig3}
  \end{figure}

\subsection{STEREO movies}

Under the column "Movies, plots, \& links" we have added a new set of daily movies displaying various combinations of images and spectra from the Solar Terrestrial Relations Observatory \citep[STEREO,][]{2008SSRv..136....5K} starting in November 2006, soon after the commissioning of the mission.  EUV and coronagraph images are from the Sun Earth Connection Coronal and Heliospheric Investigation \citep[SECCHI,][]{2008SSRv..136...67H} and the radio dynamic spectra are from the STEREO/WAVES instrument \citep{2008SSRv..136..487B}. Figure~\ref{fig2} shows list of links to the JavaScript movies. 

\subsection{Newly added CMEs}

CMEs were identified in the LASCO FOV by different people until 2004, beyond which a single person (co-author G. Micha\l{}ek) identifies CMEs. In order to make the identification uniform throughout, G. Michałek revisited the 1996--2004 images to check if any CMEs were missing. This exercise resulted in 3072 CMEs over the 9-year period. Figure~\ref{fig3} shows the statistics of the newly identified CMEs. The top panel shows the annual number of CMEs in catalog v2. The second panel from the top shows all the newly identified CMEs amounting to hundreds of CMEs in each year.  The number was high in the beginning of the mission when people were getting used to the new data set. Narrow CMEs (angular width W  $<$ 30$^\circ$) constitute the largest population among the newly added CMEs, followed by regular CMEs (30$^{\circ}\leq$ W $<$ 120$^\circ$). Only 25 new wide CMEs (W $\geq$ 120$^\circ$) were found as shown in the bottom panel. Among the wide CMEs, only one was a halo CME. 

\begin{figure}[h]
  \centering
    \includegraphics[scale=0.85
    ]{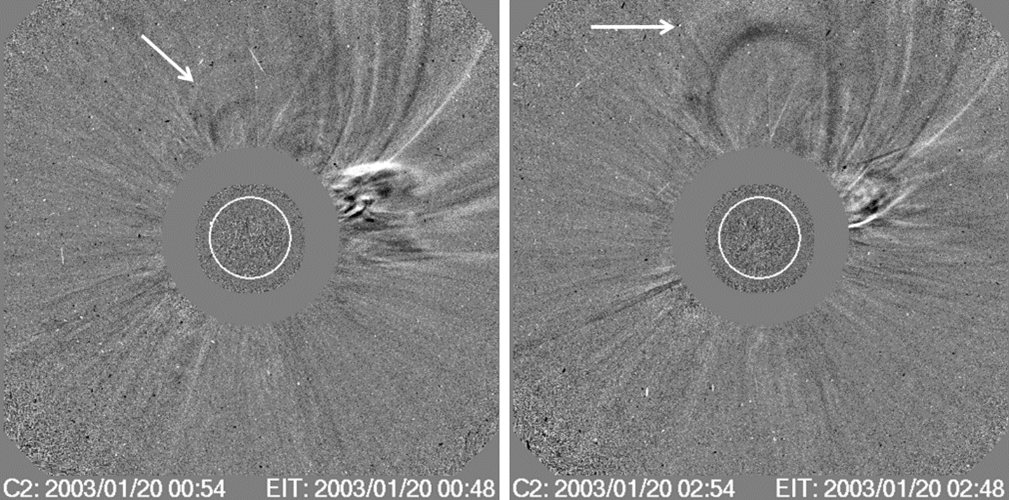}
    \caption{Running difference images of the only newly identified wide CME in the year 2003 (pointed to by the arrows). The CME is extremely faint. Dark structure behind the leading edge represents the position of the CME in the previous frame that was subtracted from the current frame to make the difference image.}
    \label{fig4}
  \end{figure}

Interestingly, most of the wide CMEs are generally faint. Figure~\ref{fig4} shows the only newly identified wide CME in the year 2003. Clearly the CME is very faint and can easily be missed. The CME is actually visible mainly in the difference images. The CME has an ID: 20030120.005405.p360g;V=233km/s indicating that the CME is also very slow. 

\begin{figure}[h]
  \centering
    \includegraphics[scale=0.35]{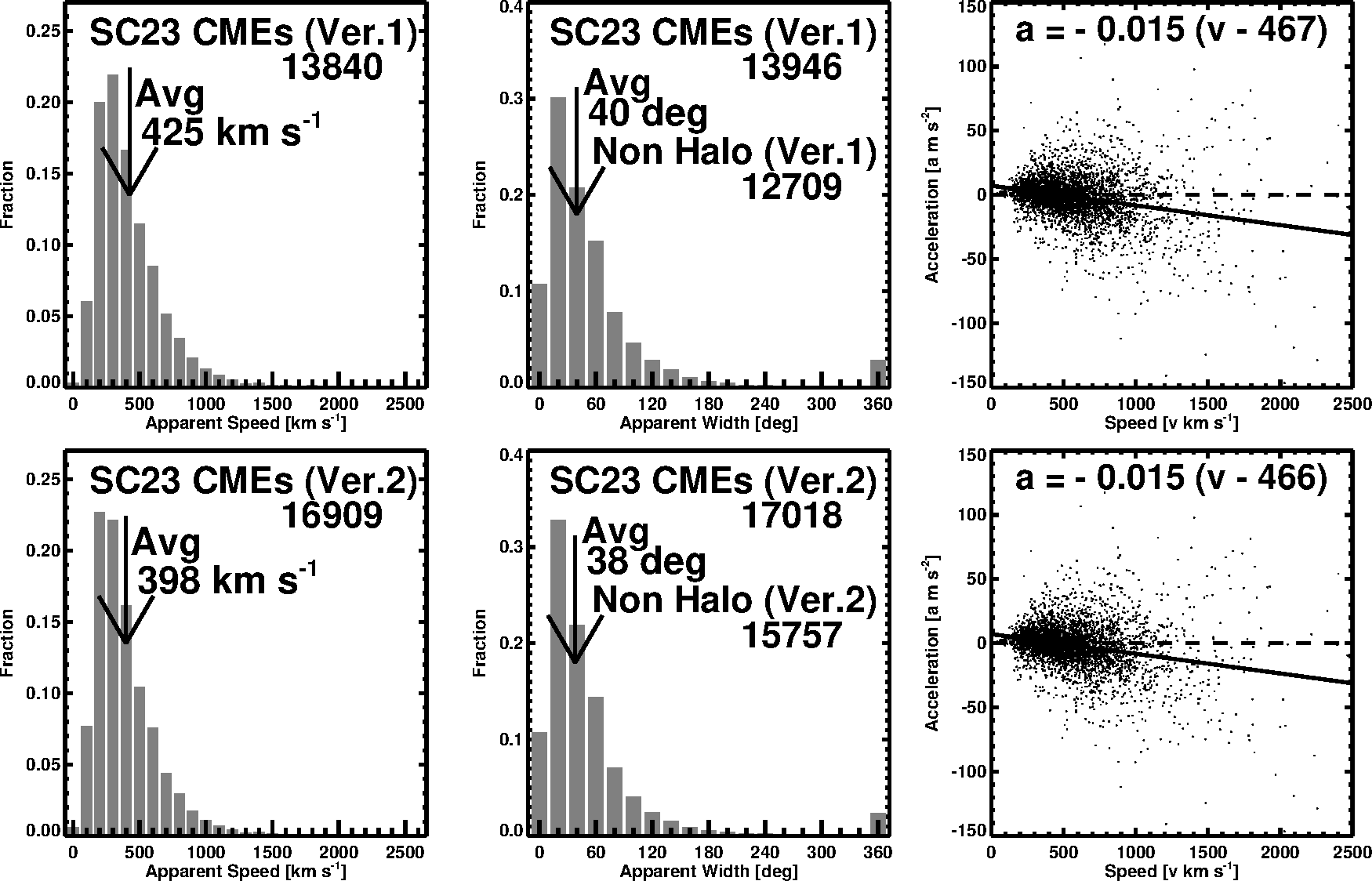}
    \caption{Speed, width, and acceleration distributions of CMEs in cycle 23 from catalog v1 (top) and the same parameters including newly added CMEs (bottom). The number of CMEs in the interval and the averages of the distributions are shown on the plots. The acceleration information is presented as a scatter plot between CME speed and the acceleration within the LASCO FOV. Note that this acceleration is typically the residual acceleration. The main acceleration peaks generally closer to the Sun (below the LASCO C2 occulting disk. Equations of the regression line is shown on the plots. E.g., a = 0.015 (v-466), where a is the acceleration in m s$^{-2}$ and v is the speed in km s$^{-1}$. Only CMEs that have five or more height-time measurements are included in the acceleration plot. CMEs with feature quality $\ge$1 are used in the speed and width distributions (i.e., ill-defined events are excluded).   }
    \label{fig5}
  \end{figure}

To see the effect of adding the new CMEs, we have compared the distributions of speed, width, and acceleration in Figure~\ref{fig5}. After the new CMEs are added, the average speed decreases by 27 km~s$^{-1}$ and the average angular width decreases by 2 degrees. The acceleration within the LASCO C2 FOV remains the same. Thus, the newly added CMEs did not affect the statistics.  

\begin{figure}[bh]
  \centering
    \includegraphics[scale=0.35]{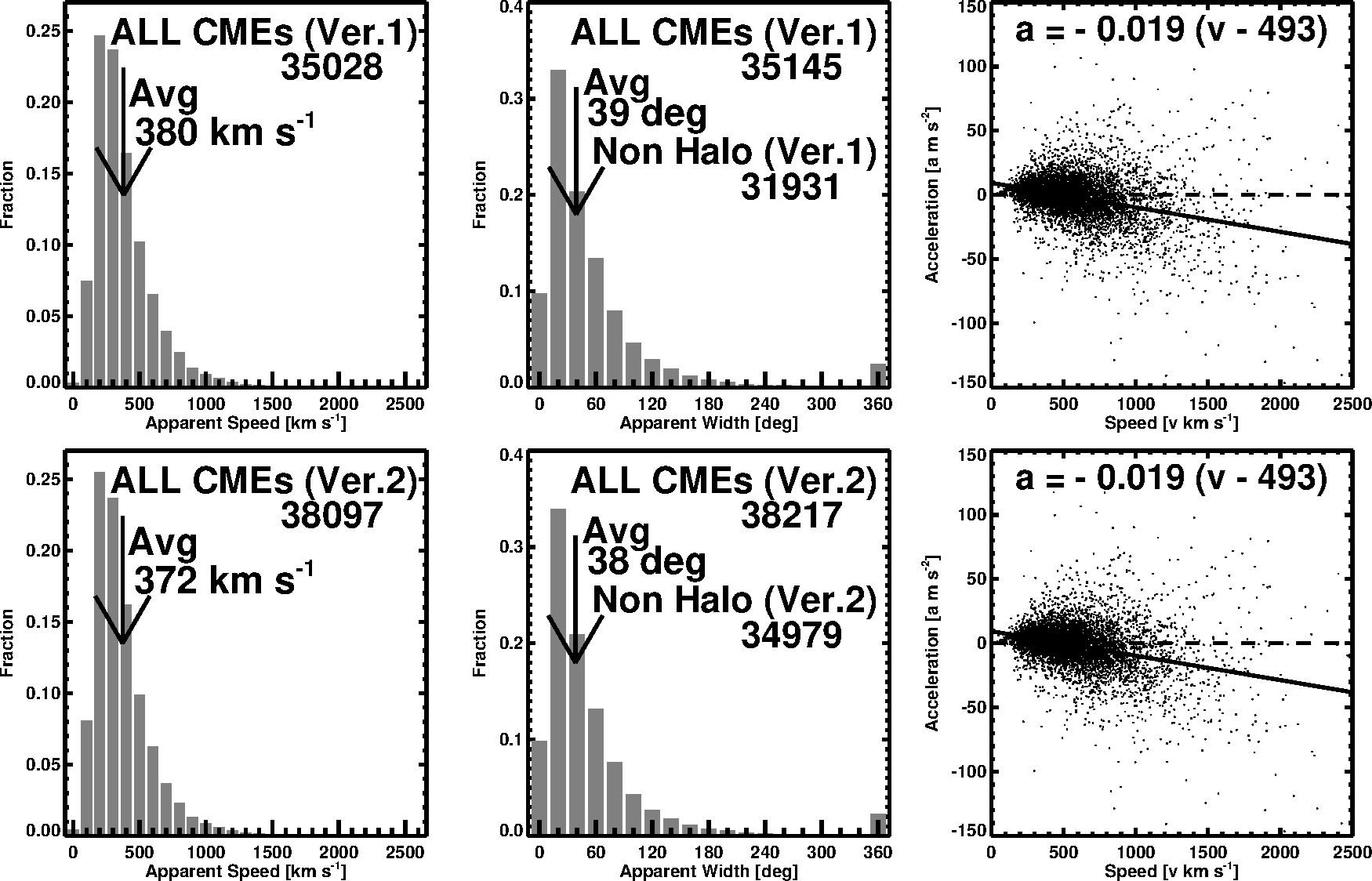}
    \caption{Same as Figure 5, but the full data sets are used (1996 to 2023, inclusive)}
    \label{fig6}
  \end{figure}

  Figure~\ref{fig6} shows distributions similar to those in Figure~\ref{fig5} except that they include all CMEs measured until the end of 2023 (includes CMEs from cycles 23, 24, and the rise phase of cycle 25). Once again, the effect of inclusion of the new CMEs is insignificant. However, we do see the effect of weak solar cycles 24 and 25: the average values are smaller because of the higher abundance of weaker CMEs in cycles 24 and 25 than in cycle 23.
\section{Summary}

As of May 1, 2024 the CDAW Data Center has released the SOHO/LASCO CME catalog v2. The new catalog has three major changes: (i) we have incorporated online tools that can be used on various images in the Javascript moes and radio dynamic spectra, (ii) we have included 29 Javascript movies that combine various images and dynamic spectra from the STEREO mission, and (iii) we revisited SOHO/LASCO data to identify missing CMEs resulting in 3070 new CMEs. Most of the newly included CMEs are weak and narrow. Inclusion of the new CMEs did not have much impact on the statistical properties of CMEs. The v1 catalog resides at
\url{https://cdaw.gsfc.nasa.gov/CME_list/index_ver1.html}
  
We thank the SOHO, STEREO, SDO, Wind, GOES, Yohkoh/SXT, and Nobeyama radioheliograph teams for their contribution in making this Catalog possible.

\bibliography{catalog}{}
\bibliographystyle{aasjournal}

\end{document}